\documentstyle[12pt]{article}

 \newcommand{\be}{\begin{equation}}
 \newcommand{\ee}{\end{equation}}
 \newcommand{\ba}{\begin{eqnarray}}
 \newcommand{\ea}{\end{eqnarray}}
 
 \newcommand{\del}{\partial}

\def\infinity{\infty}

\newcommand{\lef}{\left}

\newcommand{\ri}{\right}

\newcommand{\cl}{{\cal L}}

\newcommand{\fr}{\frac}

\begin{document}

\begin{titlepage}

\topmargin -15mm

\vskip 10mm
\vskip 40mm

\centerline{ \LARGE\bf A New Approach for Bosonization of }
\vskip 2mm
\centerline{ \LARGE\bf  Massive Thirring Model in Three Dimensions }

    \vskip 2.0cm

    \centerline{\sc R.Banerjee $^*$ and E.C.Marino }

     \vskip 0.6cm
     
\centerline{\it Instituto de F\'\i sica}
\centerline{\it Universidade Federal do Rio de Janeiro } 
\centerline{\it Cx.P. 68528, Rio de Janeiro, RJ 21945-970, Brasil} 
\vskip 2.0cm

\begin{abstract} 

We develop a new approach for bosonization based on the direct
comparison of current correlation functions and apply it to the case
of the Massive Thirring Model in three dimensions
in the weak coupling regime,
but with an arbitrary mass. Explicit bosonized forms for the
lagrangian and the current are obtained in terms of a vector gauge field.
Exact results for the corresponding expressions are also obtained in
the case of a free massive fermion. Finally, a comment on the
derivation of the
current algebra directly from the bosonized expressions is included.

\end{abstract}

\vskip 3cm
$^*$ On leave of absence from S.N.Bose National Centre for Basic
Sciences, Calcutta, India.
\vskip 3mm
Work supported in part by CNPq-Brazilian National Research Council.
     E-Mail addresses: rabin@if.ufrj.br; marino@if.ufrj.br

\end{titlepage}

\hoffset= -10mm

\leftmargin 23mm

\topmargin -8mm
\hsize 153mm
 
\baselineskip 7mm
\setcounter{page}{2}

Bosonization in spacetime dimensions higher than two has attracted a
lot of attention in the past few years \cite{em1,rb,rb1,bos,bfo}.
Even though
a direct Mandelstam-like operator bosonization
has only been found in the case of the free massless Dirac fermion field in
2+1D \cite{em1}, bosonized
expressions for the fermion lagrangian and the current have been
obtained in different mass regimes of free theories \cite{bfo}.
For interacting theories, on the other hand, only the large mass limit
has been explored \cite{rb,rb1,bos}.
General
bosonized expressions for the current and lagrangian in such interacting
theories are still lacking for an arbitrary value of the fermion mass.
Furthermore, a new question which arises in this framework and
which has not been discussed in the literature is the
issue of renormalizability of the fermionic theory when compared with
its bosonized counterpart. This is a feature which is absent in 1+1D.

In this work, we present the explicit bosonization of the Massive Thirring
Model in three dimensions (MTM) in the weak coupling regime, but for
arbitrary mass.
Contrary to previous approaches, we develop a new method of bosonization
that is based on the direct comparison of current correlation functions.
We derive explicit
bosonized forms for the current and lagrangian for arbitrary values of the
fermion mass. It is shown that the MTM for weak coupling but with
arbitrary mass is mapped into a generalized free gauge theory. This suggests
that at least in this regime we can make sense out of the MTM in spite
of the fact that it is perturbatively nonrenormalizable.
Interestingly, the bosonized version of the MTM turns out to be
nonlocal for all values of the fermion mass $m$,
except $m \rightarrow \infinity$.
We also define a dual current which has identical correlation functions
as the usual current.

We finally consider the case of a free fermion with an arbitrary mass as a
limiting situation of the MTM when the coupling constant vanishes. In this
case, we get exact results for the bosonized lagrangian and
current. In the zero mass limit these expressions reduce to the ones
obtained in \cite{em1}. In the
$m \rightarrow \infinity$ regime, we derive current algebra
relations directly from the bosonic expressions and find 
a divergent Schwinger term as a result of the infinite
mass scale.

Let us consider the euclidean generating functional of the MTM in three
dimensions,
\be
Z[J] = \int  D\psi D\bar\psi
\exp \lef\{-\int d^3z \lef [ 
\bar\psi (- \not\!\del + m ) \psi + \fr{\lambda^2}{2} j_\mu j_\mu +i
\lambda j_\mu  J_\mu \ri ] \ri \}
\label{z1}
\ee
where $j^\mu = \bar\psi \gamma^\mu \psi$ and the metric of \cite{cl}
is followed.
Eliminating the four fermion interaction in the standard way by the
introduction of an auxiliary vector field we get
\be
Z[J] = \int  D\psi D\bar\psi D A_\mu
\exp \lef\{-\int d^3z \lef [ 
\bar\psi (- \not\!\del + m +i \lambda \not\!\!\! A ) \psi
+ \fr{1}{2} A_\mu A_\mu +i
\lambda j_\mu  J_\mu \ri ] \ri \}
\label{z2}
\ee
Performing the fermion integration in the weak coupling limit, only the
two-legs one-loop graph contributes to the relevant vacuum polarization
tensor and we get
\be
Z[J] = \int   D A_\mu
\exp \lef\{-\int d^3z \lef \{ 
 \fr{ \lambda^2}{2} ( A_\mu + J_\mu ) \Pi_{\mu\nu} ( A_\nu + J_\nu )
+ \fr{1}{2} A_\mu A_\mu 
 \ri \} \ri \}
\label{z3}
\ee
where an explicit computation either on the lattice or in the continuum
yields \cite{cl,djt} in the momentum space
\be
\Pi_{\mu\nu} (q) = A(q^2) C_{\mu\nu}(q) + B(q^2) P_{\mu\nu}(q)
\label{pi}
\ee
where $C_{\mu\nu}(q) = \epsilon_{\mu\nu\alpha} q_\alpha$ and
$P_{\mu\nu}(q) = q^2 \delta_{\mu\nu}
- q_\mu q_\nu$, while
\be
A(q^2 ) = a_0 + \fr{1}{4\pi} \int_0^1 dt \{ 1 - m [ m^2 + t(1-t) q^2
]^{-1/2} \}
\label{a}
\ee
and
\be
B(q^2) = 
 \fr{1}{2\pi} \int_0^1 dt t(1-t)  [ m^2 + t(1-t) q^2 ]^{-1/2}
\label{b}
\ee
In (\ref{a}) $a_0$ is a finite coefficient depending on the
specific regularization \cite{cl, djt}. Subsequent computations
are greatly simplified by using the following algebraic identities
among $C_{\mu\nu}$ and $P_{\mu\nu}$
\be
C_{\mu\alpha}C_{\alpha\nu} = - P_{\mu\nu} \ \ ;\ \ 
P_{\mu\alpha}P_{\alpha\nu} = q^2  P_{\mu\nu} \ \ ;\ \  
C_{\mu\alpha}P_{\alpha\nu} =
P_{\mu\alpha}C_{\alpha\nu} = q^2  C_{\mu\nu}
\label{cp}
\ee

Now we can perform the quadratic functional integral over $A_\mu$, obtaining
\be
Z[J] = 
\exp \lef\{-\int d^3z  \fr{ \lambda^2}{2} \lef \{ 
   J_\mu  \Pi_{\mu\nu}  J_\nu 
- \lambda^2  J_\lambda \Pi_{\lambda\mu} \Gamma_{\mu\nu} \Pi_{\rho\nu}
   J_\rho
 \ri \} \ri \}
\label{z4}
\ee
where 
$\Gamma_{\mu\nu} = [ \delta_{\mu\nu} + \lambda^2 \Pi_{\mu\nu} ]^{-1}
  = \delta_{\mu\nu} + O(\lambda^2) $. Since we are
working up to the second order in $\lambda$, therefore, we can just
retain the $\delta_{\mu\nu}$ piece in (\ref{z4}).

It is now straightforward to evaluate the current two-point function
by taking functional derivatives of (\ref{z4}) with respect to the sources
\be
<j_\mu (q) j_\nu (-q) > = 
 \Pi_{\mu\nu}(q)
 -\lambda^2  \Pi_{\mu\alpha} (q) \Pi_{\alpha\nu} (q)
\label{jj}
\ee

Based on this observation, we can infer the form of the bosonized
lagrangean of the MTM in this limit
\be
\cl_{MTM} =  \fr{1}{2} B_\mu 
 \lef (
 \Pi_{\mu\nu}
- \lambda^2  \Pi_{\mu\alpha} \Pi_{\alpha\nu} \ri) B_\nu
\label{bl}
\ee
>From the transversality of the kernel, we can infer that
this is a gauge theory. It is nonlocal for any value of the fermion
mass except for $m \rightarrow \infinity$ as can be explicitly checked
from expressions (\ref{a}) and (\ref{b}).
Generalized free gauge theories of this type,
inspite of being nonlocal, have been studied in the literature and shown
to yield sensible results \cite{bfo,ma,bm}. The above mapping of the MTM is
one of the central results of our work.

We are now in a position to write the current bosonization formula. This is
given by
\be
j_\mu = 
 \lef (  \Pi_{\mu\nu} 
- \lambda^2  \Pi_{\mu\alpha} \Pi_{\alpha\nu} \ri) B_\nu
\label{j}
\ee
>From this bosonized expression, we can reproduce the current two point
function given by (\ref{jj}). In order to obtain this result we need the
two-point correlation funtion of the $B_\mu$-field which is given by the
inverse of the operator appearing in the quadratic lagrangean (\ref{bl}).
Of course, in order to invert such an operator, a gauge fixing term has to
be added. We also take advantage of the tranversality of $\Pi_{\mu\nu}$,
in order to add the same longitudinal
gauge piece to the second $\Pi_{\mu\nu}$
in the $\lambda$-dependent term of (\ref{bl}). The result is
$$
<B_\mu(q) B_\nu (-q) > = 
\lef [ \Pi_{\mu\alpha} (q)\lef ( \delta_{\alpha\nu} -
\lambda^2 \lef ( \Pi_{\alpha\nu} (q) +
\xi B(q^2)  q_\alpha q_\nu \ri ) \ri) +
\xi B(q^2) q_\mu q_\nu  \ri ]^{-1} 
$$
\be
= D_{\mu\nu} (q)
+ \lambda^2 \fr{P_{\mu\nu}(q)}{q^2} + O(\lambda^4)
\label{bb}
\ee
where
\vfill
\eject
$$
D_{\mu\nu}(q) =
\lef [\Pi_{\mu\nu} (q) + \xi B(q^2) q_\mu q_\nu \ri ]^{-1} 
$$
\be
= \fr{1}{q^2 [A^2 (q^2) + q^2 B^2 (q^2) ] }
\lef [ B(q^2) P_{\mu\nu} - A(q^2) C_{\mu\nu}
\ri ] + \fr{1}{\xi} \fr{q_\mu q_\nu }{q^4 B (q^2)}
\label{d}
\ee
Now, using (\ref{j}), we have
\be
<j_\mu (q) j_\nu (-q) > = 
 \lef (  \Pi_{\mu\alpha} 
- \lambda^2  \Pi_{\mu\beta} \Pi_{\beta\alpha} \ri) (q)
 \lef (  \Pi_{\nu\rho} 
- \lambda^2  \Pi_{\nu\sigma} \Pi_{\sigma\rho} \ri) (-q)
< B_\alpha (q) B_\rho (-q) >
\label{jj1}
\ee
Inserting expression (\ref{bb}) for the $B_\mu$-field correlator we
precisely reproduce equation (\ref{jj}) for the current correlation
function. It is also easy to see that all the higher correlation functions
are reproduced by the bosonic expression (\ref{j}). Observe that,
in particular, the odd functions vanish because the odd $B_\mu$-correlators
are zero in the $B_\mu$-theory.
This confirms the validity of the current bosonization formula (\ref{j})
and shows why in this limit the MTM is mapped into a generalized free
theory (\ref{bl}).

It is interesting to note that a dual current can be defined as
\be
\bar j_\mu (q) = \fr{i}{q} C_{\mu\alpha} (q) j_\alpha (q)
\label{jb}
\ee
We can easily verify from the bosonic expression (\ref{j}) 
that the dual current correlation functions are identical to those of
$j_\mu$. This generalizes a similar duality relation found in the large mass
limit \cite{rb}.

We can now address the exact bosonization of the free massive fermionic
theory. This can be obtained in the limit when the Thirring coupling
$\lambda$ vanishes. From (\ref{bl}) and (\ref{j})
we immediately obtain the bosonization
formulae for the lagrangian and current, namely
\be
\bar\psi (- i \not\! q + m ) \psi |_{\rm free}
=
  \fr{1}{2} B_\mu  \Pi_{\mu\nu}  B_\nu
\label{fb}
\ee
\be
j_\mu (q) |_{\rm free} =   \Pi_{\mu\nu} B_\nu
\label{fb1}
\ee
In the zero mass limit, the above expressions reduce precisely to
the ones found by following a direct operator bosonization of the
free massless fermion field \cite{em1}. This clearly shows that the
operator realization obtained in \cite{em1} is exact and no nonquadratic
corrections are necessary.

It is very important to emphasize that these exact bosonization formulae
are strictly valid only in the free case. In the presence of an
interaction, both expressions are modified. While the current is given by
(\ref{j}), the kinetic fermion lagrangian for the MTM, for instance,
can be computed easily from (\ref{bl}) and (\ref{j}) giving the result
\be
\bar\psi (- i \not\! q + m ) \psi |_{\rm MTM} =
\bar\psi (- i \not\! q + m ) \psi |_{\rm free} -\frac{\lambda^2}{2}
B_\sigma(q)\Bigl(  \Pi_{\sigma\lambda} \Pi_{\lambda\mu}(q)+\Pi_{\sigma
\lambda}\Pi_{\lambda\mu}(-q)\Bigr) B_\mu(-q)
\label{fb1}
\ee
This relation clearly shows the modification of the bosonization formula
for the kinetic fermion lagrangian which is produced 
when we add the Thirring interaction to the free massive lagrangian. Self-
consistency is preserved as can be inferred by setting $\lambda = 0$
in the above formula. The fact that the bosonization formulae are
changed in the presence of an interaction is a general feature of
dimensions higher than two \cite{bm}. Incidentally,
the bosonization formula (\ref{fb}) for the
free massive fermion lagrangian only was given earlier in \cite{bfo}.

Note that the bosonization method
presented here is an alternative to the usual
path integral approaches followed in \cite{rb,rb1,bos,bfo}. This is
particularly well illustrated in the bosonization of the current.
Here it is given by (\ref{j}) whereas in the path integral approaches,
discussed in the large $m$ limit \cite{rb,rb1,bos}, 
it has a purely topological structure. In a further publication \cite{bm1},
we shall prove the equivalence of these different approaches.

Let us turn now to current algebra relations. For this it is enough
to consider the free case.
As shown before, there are two distinct regimes: a nonlocal theory
for any finite mass but a local one for $m \rightarrow \infty$.
These two cases must be considered distinctly.
Let us first examine the large
mass limit \cite{rb} in which the theory given by (\ref{bl}) reduces to
the Maxwell-Chern-Simons lagrangian 
\be
\cl_{MTM} = - \fr{1}{2} a_0 \epsilon_{\mu\nu\alpha} 
B_\mu \del_\nu B_\alpha + \fr{1}{48 \pi m} B_{\mu\nu} B_{\mu\nu}
\label{bl1}
\ee
where $B_{\mu\nu}=\del_\mu B_\nu -\del_\nu B_\mu$. The bosonized current
in this limit is obtained from (\ref{j}), namely
\be
j_\mu =  a_0 \epsilon_{\mu\alpha\nu} \del_\nu B_\alpha
\label{j1}
\ee
Now, the only nontrivial commutator can be explicitly evaluated. It just
requires the canonical commutator algebra between the electric and
magnetic fields in Maxwell-Chern-Simons theory \cite{djt}. We find,
in Minkowski space,
\be
[j^0 (\vec x, t), j^i (\vec y, t) ] =
12 \pi \ m\  a_0^2 \  \del^i \delta^2 ( \vec x -\vec y)
\label{com}
\ee
which produces, in the $m \rightarrow \infty$ limit,
the well known divergent Schwinger term, obtained by point splitting or BJL
approaches in the fermionic version \cite{ad}.
It is interesting to see that the mass
parameter plays a role analogous to the
inverse point splitting regulator in this limit. To discuss the current
algebra for a finite mass, a complete canonical analysis of the nonlocal
bosonized lagrangian is required, which is beyond the scope of the
present paper. However, since the Schwinger term in the
current algebra should be independent
of the mass, it expected that 
the nonlocality of the bosonized current and
lagrangian is probably responsible for yielding this term.
This will be explored in a future work.

We conclude by stressing the practical nature of
this new approach to bosonization
in higher dimensions, which is just
based on the comparison of current correlation functions. Using these ideas
the bosonization of the MTM was performed in the weak coupling
regime. Explicit expressions for the lagrangian and current were
derived for an arbitrary mass. This was also exploited to obtain the
exact bosonization formulas for a free fermion for any value of the mass.
Another important result which completely departs from the usual 1+1D case
is the demonstration that operator bosonization formulas in general depend on
the theory in which these operators are embedded.
In all cases the bosonized expressions were given solely in terms of the
vacuum polarization tensor $\Pi_{\mu\nu}$. This  general structure
suggests the possibility of using our technique in other higher
dimensions. We report on this and related aspects in a future work
\cite{bm1}.
Finally, let us remark
that the finiteness of the vacuum polarization tensor $\Pi_{\mu\nu}$
allows us to conclude that at least in the small coupling regime one
can make sense out of the MTM which is otherwise nonrenormalizable.

\vfill\eject

\leftline{\Large\bf Acknowledgements} \bigskip

Both authors  
were partially supported by CNPq-Brazilian National Research Council.
RB is very grateful to the Instituto de F\'\i sica-UFRJ 
for the kind hospitality.

\end{document}